\documentclass[pre,twocolumn,superscriptaddress,showpacs,amssymb]{revtex4}
\usepackage{amsmath}
\usepackage[utf8x]{inputenc}
\usepackage[dvipdfm]{graphicx}
\usepackage[german,english]{babel}

\begin{document}
\title{Phase Diagram for a 2-D Two-Temperature Diffusive XY Model}
\author{Matthew D. \surname{Reichl}}
\author{Charo I. \surname{Del Genio}\footnote{Present address: \foreignlanguage{german}{Max Planck Institut für Physik complexer Systeme, Nöthnitzerstraße 38, 01187 Dresden, Germany.}}}
\author{Kevin E. \surname{Bassler}}
	\affiliation{\mbox{Department of Physics, 617 Science and Research 1, University of Houston, Houston, Texas 77204-5005, USA} \mbox{and Texas Center for Superconductivity, 202 Houston Science Center, University of Houston, Houston, Texas 77204-5002, USA}}

\date{\today}

\begin{abstract}
Using Monte Carlo simulations, we determine the phase diagram of a diffusive two-temperature
conserved order parameter XY model. When the two temperatures are equal the system becomes
the equilibrium XY model with the continuous Kosterlitz-Thouless (KT) vortex-antivortex unbinding
phase transition. When the two temperatures are unequal the system is driven by an energy flow
from the higher temperature heat-bath to the lower temperature one and reaches a far-from-equilibrium
steady state. We show that the nonequilibrium phase diagram contains three phases: A homogenous
disordered phase and two phases with long range, spin texture order. 
Two critical lines, representing
continuous phase transitions from a homogenous disordered phase to two phases of long range
order, meet at the equilibrium KT point. The shape of the nonequilibrium critical lines as they
approach the KT point is described by a crossover exponent $\varphi = 2.52 \pm 0.05$. Finally,
we suggest that the transition between the two phases with long-range order is first-order,
making the KT-point where all three phases meet a bicritical point.
\end{abstract}

\pacs{05.70.Ln 64.60.Kw 64.60.F- 64.60.Cn}

\maketitle

Much of the research in the statistical physics of nonequilibrium sytems
has been directed toward understanding how universal equilibrium critical
phenomena are affected by dynamical nonequilibrium perturbations. Field-theoretical
studies have indicated that the effects of nonequilibrium dynamics are drastic
in systems where detailed balance violation is coupled with conserved anisotropic
dynamics~\cite{Sch95}. In these systems, effective long range interactions
can be induced by the \emph{local} dynamics producing a critical behavior
that is remarkably different from the one of the corresponding unperturbed,
\emph{equilibrium}, systems~\cite{Kat83,Kat84,Gri85,Jan86,Leu86,Sch93,Bas94a,Bas94b,Bas94c,Bas95a,Bas95b,Sch96,Tau97,Sch98,Odo04,Ris05,Zia10}.

In this paper, we present the phase diagram for a two-dimensional two-temperature diffusive
conserved order parameter XY model. The system evolves through 
Kawasaki spin-exchange dynamics~\cite{Kaw66}. Thus, the dynamics
is purely relaxational with no reversible mode couplings, 
and corresponds to Model~B of Ref.~\onlinecite{Hoh77}.
Long range order can exist in nonequilibrium steady states of this system due to the effective
long range interactions generated by the anisotropic diffusive dynamics that occurs in that
regime. The ordered phase is characterized by the appearance of standing spin waves, or spin
textures, oriented along the direction of lower temperature. The system exhibits a nonequilibrium
disorder--long-range order transition that is in the same universality class as an equilibrium
model with dipole interactions~\cite{Bas95a,Tau97}. Note that our model reduces to the \emph{equilibrium}
XY model in the limit where both temperatures are equal. Also, the Mermin-Wagner theorem states
that there is no spontaneous symmetry breaking in equilibrium systems with continuous symmetry
of the order parameter and dimension $d=2$~\cite{Mer66}. Thus, no long-range ordered phase
is observed in the two-dimensional equilibrium XY model. However, the equilibrium system still
undergoes a transition from quasi-long-range order to disorder characterized by the emergence
and unbinding of vortices and antivortices, which is the Kosterlitz-Thouless~(KT) transition~\cite{Kos73}.
Hereafter we refer to the quasi-long-range order phase as the KT phase. Since both a KT
transition and a disorder--long-range order phase transition occur in the two-dimensional two-temperature
XY model, we expect to find a KT--dipole crossover in the phase diagram for this system.

Using results from Monte~Carlo simulations, we show that two critical lines representing nonequilibrium
disorder--long-range order transition temperatures meet at the equilibrium KT transition temperature.
These lines are described by an exponent which we predict to be the universal exponent for KT--dipole
crossover. Finally, we argue that, at temperatures below the critical KT temperature, any infinitesimal
nonequilibrium perturbation to the system, will produce long-range ordered phases. Thus, the nonequilibrium
behavior is very different than that in equilibrium where long-range order is forbidden due 
to the Mermin-Wagner
theorem~\cite{Bas95a}.

Our model consists of a set of two-dimensional spins arranged on a square lattice of rectangular
dimensions $L_x$ and $L_y$. Each spin $\vec s_i$ is a unit vector. The directions of the spins
are evenly distributed from $0$ to $2\pi$ over the lattice, so that their vector sum is null. The
total energy of the system is given by the Hamiltonian
\begin{equation*}
 H=-\sum_{\left\langle ij\right\rangle}\vec s_i\cdot\vec s_j\:,
\end{equation*}
where $\left\langle ij\right\rangle$ indicates sum over the nearest neighbor spins on the lattice.
The system evolves through Kawasaki exchanges with Metropolis rates~\cite{Kaw66,Met53}. The exchanges
along the $x$ and $y$ axes satisfy detailed balance with temperatures $1/\beta_x$ and $1/\beta_y$,
respectively. When $\beta_x\neq\beta_y$, an energy current flows from the hotter heat bath to the
cooler one and detailed balance is no longer satisfied globally. When this is the case, phase transitions
occur in nonequilibrium steady states and are characterized by the appearance of a long-wavelength
spin texture in the direction with the larger value of $\beta$. In our nonequilbrium simulations,
we generally study the case $\beta_y < \beta_x$, so that the spin texture appears in the $x$ direction.
To give a quantitative measure of this ordering, we define the order parameter $\Psi$ as the ensemble
averaged arithmetic average of the components of the long-wavelength limit of the structure factor:
\begin{equation*}
 \Psi = \frac{1}{2}\left[C_1\left(\frac{2\pi}{L_x}, 0\right)+C_2\left(\frac{2\pi}{L_x}, 0\right)\right]\:,
\end{equation*}
where $C_n\left(k_x, k_y\right)$ is the normalized Fourier transform of the $n^\mathrm{th}$ component
of the spin vectors of our system.

The spatial anisotropy of the system requires an analysis using anisotropic finite
size scaling~\cite{Leu91}. Hence, one must compare systems with sizes that scale in
a way that keeps the expression $L_x^{1+\Delta}/L_y$ constant, where $\Delta$ is the
anisotropy exponent. The value $\Delta = 1$ has been estimated using renormalization
group techniques to first order in a dimensional epsilon expansion~\cite{Tau99}. Therefore,
we performed simulations on systems of sizes $12\times9$, $16\times16$, $24\times36$
and~$32\times64$. Note that there may be higher-order corrections to the value of
$\Delta$ that, with this choice of system sizes, would introduce some systematic errors
in the data analysis.
\begin{figure}
 \centering
\includegraphics[width=0.45\textwidth]{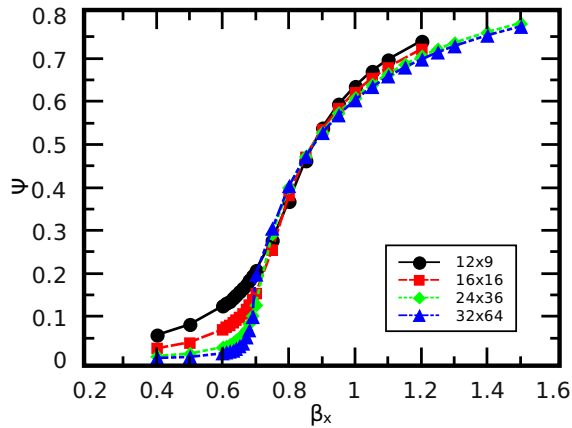}
\caption{\label{ordpar}(Color online). Order parameter $\Psi$ vs.\ $\beta_x$ at $\beta_y=0$ for different
system sizes. The black circles and solid line correspond to a system size of $12\times 9$;
the red squares and dashed line correspond to a system size of $16\times 16$; the green diamonds
and dotted line correspond to a system size of $24\times 36$; the blue triangles and dashed-dotted
line correspond to a system size of $32\times 64$. The plot clearly shows a transition from
disorder to order at a $\beta_x$ of approximately $0.7$. The error bars are smaller than
the symbol size.}
\end{figure}
\begin{figure}
 \centering
\includegraphics[width=0.45\textwidth]{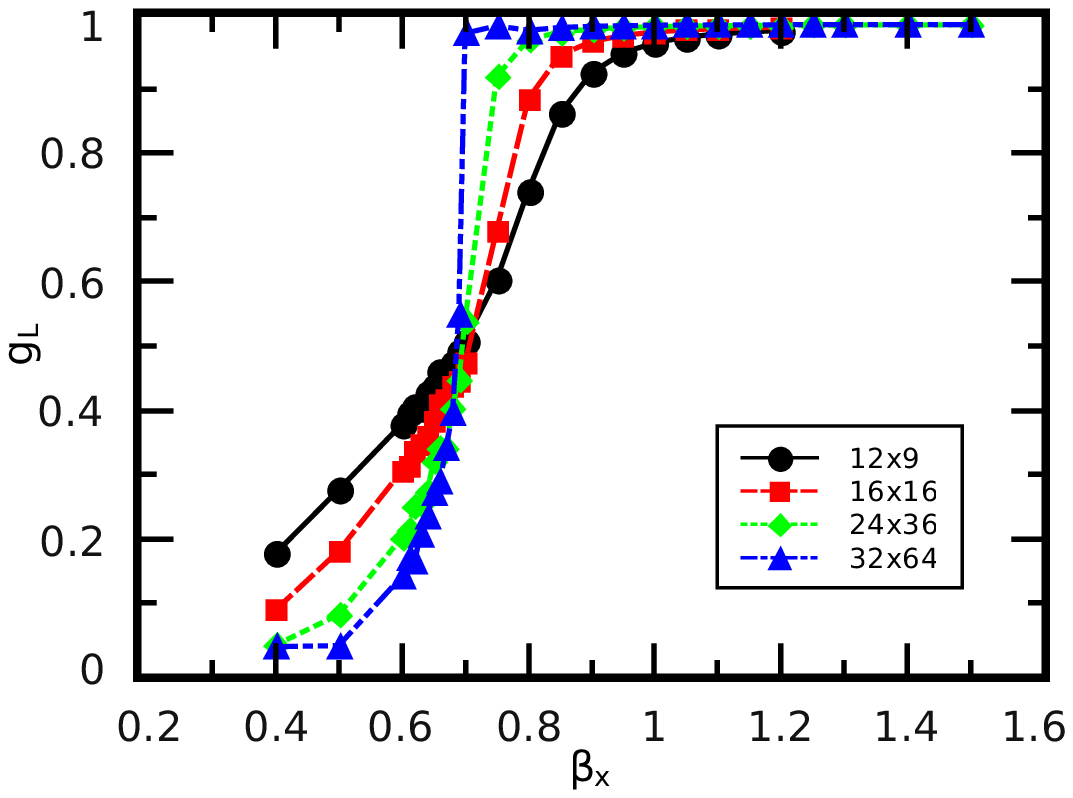}
\caption{\label{bincum}(Color online). Binder's cumulant $g_{L}$ vs.\ $\beta_x$ at $\beta_y=0$ for different
system sizes. The black circles and solid line correspond to a system size of $12\times 9$;
the red squares and dashed line correspond to a system size of $16\times 16$; the green diamonds
and dotted line correspond to a system size of $24\times 36$; the blue triangles and dashed-dotted
line correspond to a system size of $32\times 64$. The plots intersect asympotically at $\beta_x\approx0.68$.
The error bars are smaller than the symbol size.}
\end{figure}

We measured $\Psi$ after each Monte Carlo sweep~(MCS) during the simulations.
After estimating the relaxation time, we determined the ensemble averages $\langle\Psi\rangle$
and $\langle\Psi^2\rangle$ over uncorrelated configurations in the steady state.
We ran $4\times10^6$, $5\times10^6$, $8\times10^7$ and $10\times10^8$ MCS for
the system sizes $12\times9$, $16\times16$, $24\times36$ and $32\times64$, respectively.
Integrated autocorrelation times ranged from roughly 200~MCS for the smallest
system to roughly 1200~MCS for the largest system.

Our simulations reveal that long-range ordered states occur when $\beta_y$ is sufficiently
small and $\beta_x$ is sufficiently large or viceversa, by symmetry. Note that the $\beta_x$--$\beta_y$
phase diagram for this system is symmetric about the diagonal since interchanging these temperatures
is equivalent to simply renaming the axes of the lattice. Thus, we study the ordering process
only in the $\beta_y<\beta_x$ region.

Figure~\ref{ordpar} shows the value of $\Psi$ as a function of $\beta_x$ with $\beta_y=0$
for different system sizes. The data clearly show ordering occuring at $\beta_x\approx0.7$.
Graphs showing similar critical behavior were produced for each simulated value of $\beta_y$.
We achieved more precise estimates of the disorder--order transition temperatures by measuring
the crossing point of Binder's cumulant
$g_L\equiv 3-2\left(\left\langle\Psi^2\right\rangle/{\langle\Psi\rangle}^2\right)$~\cite{Lan00}.
As expected for a continuous phase transition, the values of $g_L$ for different system sizes
cross at the critical point $\beta_{x_\mathrm c}$, as shown in Fig.~\ref{bincum}. This allowed
us to measure the critical $\beta_x$ for $\beta_y$ values of $-0.9$, $-0.75$, $-0.6$, $-0.3$,
0, $0.3$, $0.6$, $0.75$, $0.9$ and 1.
\begin{figure}
 \centering
\includegraphics[width=0.45\textwidth]{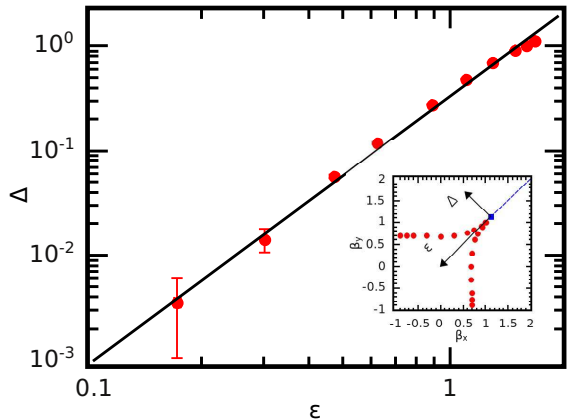}
\caption{\label{LL}(Color online). Power-law fit of the dependence of $\delta$ on $\varepsilon$
for the measured critical points. The points lying on a straight line on a double logarithmic plot
imply $\delta\sim\varepsilon^\varphi$. The slope of the line corresponds to the universal crossover
exponent $\varphi$. We measure $\varphi=2.52\pm 0.05$. The inset shows the geometric representation
of the $\varepsilon-\delta$ coordinates.}
\end{figure}

The locations of the transition points can be parametrized in terms of the
quantities $\varepsilon = \left(2\beta_{KT}-\beta_{x_c}-\beta_{y_c}\right)/\sqrt{2}$
and $\delta=\left(\beta_{x_c}-\beta_{y_c}\right)/\sqrt{2}$.
The result of such parametrization is shown in a log-log plot in Fig.~\ref{LL}.
The possibility of fitting the data to a straight line implies they obey the
power law $\delta\sim\varepsilon^\varphi$. The slope of this line allows us
to estimate the crossover exponent as $\varphi=2.52\pm0.05$.

\begin{figure}
 \centering
\includegraphics[width=0.45\textwidth]{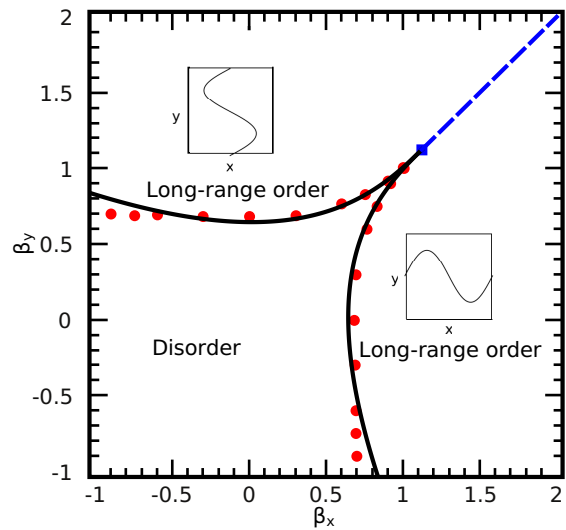}
\caption{\label{betafit}(Color online). Phase diagram for the two-dimensional two-temperature XY model. The red points
are actual measurements of the nonequilibrium transition temperatures between disorder and long-range
order phases. The two solid black critical lines are drawn with a crossover exponent $\varphi=2.52$ to fit the
points. The blue dashed line corresponds to the equilibrium KT phase. The blue square corresponds to the
KT transition temperature. The insets show schematic drawings of spin textures in the ordered state. The
error bars are smaller than the symbol size.}
\end{figure}

Figure~\ref{betafit} shows the phase diagram for the model. The insets in the figure
show the alignment of the spin textures associated with ordered states. The two critical
lines representing the nonequilibrium disorder--long range order transitions meet at
a temperature $1/\beta_c\approx0.89$. This is the same as the KT critical temperature~\cite{Gup92},
leading us to conclude that the ordered regions of the diagram meet at a line corresponding\
to the low temperature equilibrium KT phase.

Monte~Carlo simulations were also used to investigate the low temperature behavior of our system
near equilibrium. We measured the time evolution of the difference $\delta\Psi$ between the order
parameters for the $x$ and $y$ directions. This was done while varying $\beta_x$ and $\beta_y$ so
that, as time (MCS) progressed, we moved perpendicularly across the equilibrium line from one long-range
ordered region to the other. Hysteresis was clearly observed for square systems smaller than $32\times32$,
indicating that the equilibrium line may also be a line of first order transitions
from order in one direction to order in the other.

However, for larger system sizes, a different, glassy-type of behavior was observed.
The quantity $\delta\Psi$ did not switch from a large positive (negative) value to a
large negative (positive) value, as in a hysteresis loop, but stayed at a value of
approximately 0 after crossing the equilibrium line. The actual system configurations
where $\delta\Psi\approx0$ were investigated, and we observed columns of ordered vectors,
pointing in roughly the same direction, between columns of disordered vectors.

We note that a similar ``striped'' configuration was observed after deep quenches
from the high temperature disordered phase to a low temperature ordered phase. In
the steady state simulations described above, we avoided these striped configurations
in favor of long wavelength spin textures by starting the simulations in the steady
state configuration of a temperature very close to the one currently being simulated;
this process was continued for monotonically decreasing temperatures from the disordered
phase to the ordered phase. Thus, quenching of the system to a striped configuration
was avoided.

We believe the striped configuration to be a metastable state that can be found in
finite-sized systems with anisotropic nonequilibrium dynamics, e.g.\ a driven Ising
model~\cite{Hur02}. We caution however that in the case of a driven diffusive Ising
model, such striped configurations are \emph{stable} in the thermodynamic limit $L\rightarrow\infty$~\cite{Lev01,Zia00}.
In particular, if $L_\parallel$ is the dimension of the drive and $L_\perp$ the other
dimension, ``wide'' systems ($L_\perp>L_\parallel$) and square systems
support stable striped configurations, while ``narrow'' systems ($L_\parallel\gg L_\perp$)
support stable long-range ordered configurations (a single stripe in the case of the
Ising model). In light of these results for the driven Ising model, we note that it
is possible that the striped configurations in our model may represent true stable
states, particularly for the large \emph{square} systems used in the simulations
exploring the low-temperature region.

In any case, these striped configurations do not indicate an extended KT phase
in the low temperature region since the KT phase, characterized by the appearance
of vortices, is entirely different from the long-range ordered phase or the striped
phase. Thus, the phase diagram is drawn to indicate that infinitesimal non-equilibrium
perturbations to the dynamics of the system cause completely different system
behavior at temperatures below the KT transition temperature.

This is consistent with what can be inferred from the following argument.
Consider the Langevin equation for model B with a two-component order
parameter $\vec{\eta}$:
\begin{equation*}
 \partial_t\vec\eta=\lambda\nabla^2\left[\left(-\nabla^2+\tau\right)\vec\eta+\frac{1}{6}g\vec\eta\eta^2\right]+\vec\xi\:,
\end{equation*}
where $\vec\delta$ is the order parameter field, $g$ and $\lambda$ are generic constants,
$\vec\xi$ is a Gaussian noise term and $\tau\equiv\frac{T-T_c}{T_c}$ is the reduced temperature.
To account for the system having two different temperatures, the operators, the parameters
and the noise term are split into $x$ and $y$ components:
\begin{equation*}
\begin{split}
 \partial_t\vec\eta=&\lambda\left\lbrace\partial_x^2\left[\left(-\partial_x^2+\tau_x\right)\vec\eta+\frac{1}{6}g_x\vec\eta\eta^2\right]+\right.\\
&\left.\partial_y^2\left[\left(-\partial_y^2+\tau_y\right)\vec\eta+\frac{1}{6}g_y\vec\eta\eta^2\right]\right\rbrace+\\
&\xi_x\hat x+\xi_y \hat y+2\partial_x^2\partial_y^2\vec\eta\:,
\end{split}
\end{equation*}
where $\hat x$ and $\hat y$ are the unit vectors in the $x$ and $y$ directions, respectively.
To describe the system below criticality close to the equilibrium line, both $\tau$s should
be negative~\cite{Sch95}. This means that the system is unstable with respect to perturbations
in both directions. The case $\tau_x=\tau_y$ corresponds to the equilibrium model, in which the
instability has the same importance in both directions. However, any nonequilibrium perturbation
will cause the instability to become stronger in one of the two directions, thus generating effective
long range correlations in the direction corresponding to the lower $\tau$. The possibility of
an extended KT phase is ruled out by the established result that, for large enough systems, the
contribution to the correlation between far away spins due to vortices is vanishingly small~\cite{Kos74}.
This means that any long range interaction, however small, is enough to make vortices unimportant
in the description of the system. Consequently, the equilibrium KT phase is destroyed by any
nonequilibrium perturbations.

In conclusion, we determined the phase diagram for a two-dimensional two-temperature
conserved order parameter XY model. The system, whose evolution
happens through Kawasaki spin exchanges, has lines of continuous phase transition between
ordered states, characterized by the appearance of spin textures in the direction of
the lower temperature, and a disordered state. These lines meet at the equilibrium KT
point, corresponding to a KT transition in the equilibrium model. We measured the crossover
universal critical exponent $\varphi$ for this transition, finding the value $\varphi=2.52\pm0.05$.
For temperatures lower than the KT point, the equilibrium line on the phase diagram
is a line of first order transition between the ordered states, with the direction of
the spin textures changing. We provided an argument validating our finding and excluding
the possibility of an extended KT phase below criticality.

\begin{acknowledgments}
This work was supported by the NSF through Grant No.~DMR-0908286 and by
the Texas Center for Superconductivity at the University of Houston (T${}_\mathrm{c}$SUH).
\end{acknowledgments}

\end{document}